# Does Broadband Connectivity and Social networking sites build and maintain social capital in rural communities?

**Sanjib Tiwari**
School of Management and Enterprises
University of Southern Queensland
Queensland, Australia
Email: sanjib.tiwari@usq.edu.au

**Michael Lane**
School of Management and Enterprises
University of Southern Queensland
Queensland, Australia
Email: michael.lane@usq.edu.au

**Khorshed Alam**
School of Commerce
University of Southern Queensland
Queensland, Australia
Email: khorshed.alam@usq.edu.au

## Abstract

Broadband is a general purpose technology and major enabler for building social capital (SC) by better connecting rural communities – locally, nationally and internationally with families and friends and broadening their circles of influence. The main objectives of this paper are to determine to what extent broadband connectivity and Social Networking Sites (SNS) can facilitate building and maintaining SC in rural households. A large scale survey collected empirical data in the Western Downs Region of Queensland, Australia regarding households' adoption and use of broadband Internet including SNSs and the contribution to building SC in rural communities. The results of this study suggest that Broadband connectivity would appear to build and maintain two dimensions of SC, namely bonding and bridging for households in rural communities in the study area. Moreover SNS users appeared to have significantly higher levels of SC than non-SNS users in rural communities with Broadband connectivity.

## 1   INTRODUCTION

The growing popularity of social media has created new ways of collaboration and communication in our society. Social Network Sites (SNS) such as Facebook, Twitter, LinkedIn, Instagram and others are types of online virtual communities, that play host to hundreds of millions of users and have gained popularity since early 2000s (Hu and Ma 2010). The main purpose of joining any SNS is to connect and share with other users by creating a personal network (Cheung and Lee 2010). Increasingly, people have integrated their daily routines into SNS and share their interests and activities, with family, relatives and friends by posting information and events in SNS (Shin 2010). These activities allow SNS users to engage in social activities and build and maintain social network in online and offline settings among family and friends (Ellison et al. 2006). Growing popularity of social networking sites has created a new stream of inquiry for academics and practitioners alike, as indicated by the number of published works relating to online social networks and social well-being in both organizations and communities (Lo and Riemenschneider 2010).

Recent studies show that Information and Communication Technologies (ICTs) play an important role in building the national economy and social well-being (Ahmed and Al-Roubaie 2013; Chakraborty and Mandal 2014; Revi and Rosenzweig 2013). There is also growing global recognition of the discrete benefits of broadband Internet connectivity and the use of social media as a mechanism to overcome rural disadvantage and promote social well-being (Warburton et al. 2013). Once connected, rural communities can benefit from the applications and services that can be accessed via the Internet such as communication, telemedicine, distance education, e-commerce, and telework (Stenberg et al. 2009). More importantly, access to broadband technologies could help strengthen social networks with family, relatives and friends. However, Simpson (2005) citing Grunwald (1997) notes that technology initiatives alone are not a panacea unless they are embedded in the process of community development. The emphasis is not on the technology and what it can do, but on how the technology can be used strategically to meet community needs. The intent of study then should extend beyond solving the 'divide' through mere access to broadband technology. Instead focus should be on strategies that foster social inclusion, mobilise community support for achieving community goals, and thereby 'multiply' the existing community assets (Warschauer 2003). However, most study focus on building social capital among younger people in school, universities and particularly in urban locations (see Ahn 2012; Chowdhury et al. 2012; Ellison et al. 2006). There is little empirical research on the extent to which broadband connectivity and social networking sites facilitates building and maintaining Social Capital in rural communities.

The purpose of this paper is to determine to what extent Broadband connectivity and SNS can build SC for households in rural communities. Choosing rural communities in the Western Downs Region as a case study is interesting and important because of the digital divide and geographical isolation that is more likely to exist in rural areas in comparison to urban areas (ABS 2014A). Moreover, it is topical in political discourse that rural development can be enhanced by local factors of cohesion and identity and broadband connectivity may play a key role (Callois and Aubert 2007) in building and maintaining this social cohesion through social capital. This paper is organized as follows. Firstly, social capital is defined and discussed in the context of Broadband Internet access. Then SNSs are defined and the role of SNS in building SC in rural communities is discussed. Then research methodology is described and justified. The key findings regarding the results of the data analysis are discussed. Finally this paper concludes by describing the major contributions for research and practice, the limitations of this research and suggests areas for future research.

## 2 LITERATURE REVIEW

### 2.1 Social Capital

Despite the lack of a consensus on a precise definition, the term 'social capital' is extensively accepted and used as a multidimensional concept (Warburton et al. 2013). An extensive review of literature shows that researchers have defined the construct of SC in terms of social networks, trust, civic engagement, life satisfaction and other concepts (Bourdieu 1985; Coleman 1988; Lin 2002; Putnam 2001). Putnam (2001, p. 19) compares SC to the "connecting among individuals – and social networks and the norms of reciprocity and trustworthiness that arise from them". Drawing on the theory developed by Coleman (1988) and Bourdieu (1985), SC is considered as a resource that may be used when it is shared to achieve a variety of ends. Moreover, the basic idea of social capital is simple as the resource available to access and usage by individuals or groups of people through social interactions and communication among communities.

To build an understanding of the discrete processes and foundations of SC, it is essential to consider two key elements of SC: bonding and bridging SC. Bonding describes, SC generated and shared by members of a relatively homogenous group, in terms of the strong or close ties of similar groups of people founded by shared values, accepted thoughts and social norms, such as families, relatives, friends or neighborhood groups (Warburton et al. 2013; Woodhouse 2006). The resources that are available through one's strong ties correspond to bonding capital. Strong ties tend to be the source of primary personal interaction and support (Hampton 2011; Haythornthwaite 2005; Straits 2000). Bonding SC provides personal, social and emotional support (e.g. look after someone when they feel not good or sick), which plays a role in maintaining close relations (Lin 2002). On the other hand, bridging refers to SC generated and shared through interconnections between heterogeneous groups and more diverse, i.e. weak ties. Weak ties are more crosscutting than strong ties and present a lower level of homophile when compared with strong ties (Hampton 2011). People with weak ties have access to different resources, such as information and job leads that would be otherwise unavailable to them through their close ties (Granovetter 1995; 1973). So, bridging SC allows individuals to access resources not available in their close social networks. Bridging capital is useful to gain resources, i.e. for instrumental actions such as finding a job (Lin 2002). Bridging SC draws on outside or peripheral knowledge, resources and ideas that can help communities interconnect with other communities. The concepts of bonding and bridging SC are based on similar norms of trust and capacity to build the network or groups to be connected in rich social networks (Warburton et al. 2013; Woodhouse 2006). The term generalized social capital relates to a generalized trust in and reciprocity with other people (including strangers) in the wider society. Generalized trust and reciprocity is an extension of bonding social capital (Putnam 2001), and according to him, generalized trust is strongly related to other forms of civic engagement SC. Chong et al. (2011) also argues that SC among neighbors is related to generalized trust.

SC in its two key dimensions, bonding capital and bridging capital, is found across a diversity of settings from informal to the most formal social arrangements. However in defining SC as a multidimensional concept, it is obvious that SC is intangible and cannot be seen or touched. As such SC cannot be formed in isolation and instead is the product of people's association and communication with others. Individuals with a large and diverse network of contacts are thought to have more SC than individuals with small or less diverse networks. Although people often generate SC as a result of their daily communication and interaction with friends, colleague and outsiders, people also potentially make a mindful investment in social interaction (Valenzuela et al. 2009). This is the reason why many people join a SNS.

## 2.2　Social Capital and Rural communities

Rural areas are left behind in terms of developments even though these are major source of export earnings, along with the agricultural and resources sectors (ABARES 2011). A number of factors such as income, living cost, and poor access to education and health indicate that there are significant disadvantages to living in a rural area in comparison to an urban area (Cheers 1998). These issues are a source of disadvantage which make younger generations more likely to migrate to cities for better access to higher education and related employment opportunities. Increasingly there is inadequate provision of federal and state government services in rural areas. In response non-government and religious organizations, volunteer organisations such as Neighborhood Centres and local government need to work together and help each other to overcome any difficulties they face in delivering community services in rural areas (Alston 2002). However, previous research show that people who live in rural communities usually have good relationships and there is higher levels of trust because they know all or most neighbors in their community (Centre 2010; Onyx and Bullen 2000). Most of the people in these rural communities participate in various community groups with a common interest such as art, craft, community support and sporting groups (Alston 2002). Communities are greater than the sum of their parts. Rather than simply an aggregate of individuals, communities are characterized by the relationships, networks, activities, and functions that the individuals create and build together. These rural communities through working together in groups help to build strong connections and social capital among them. Previous studies shows that rural communities are founded on strong social capital at the community level (Hofferth and Iceland 1998).

## 2.3　Broadband connectivity and Social Capital in rural communities

People are increasingly reliant on the Internet and other ICTs in their business, academic, and personal spheres. The Internet is a pervasive medium through which individuals can engage in everything from personal communication through to civic participation. The Internet can serve as a vehicle for communication on formal (e.g., professional communication) and informal (e.g., emailing friends and family members) levels, as well as a source for entertainment and social activities (Quan-Haase and Wellman 2004). Broadband Internet provides a higher speeds of data transmission and is facilitating the digitalization of the economy and society in general. Broadband Internet connectivity has enabled the emergence of new work practices, home-based entrepreneurship and job searches (Autor 2001; Krueger 2000; Fairlie 2006; Stevenson 2008). More importantly, Broadband Internet connectivity can provide potential benefits in finances, health, education, entertainment, social activities and politics that can improve one's life chances (LaRose et al. 2007). Because people can use Broadband Internet to engage socially and civically, the technology is recognized as an important tool for many different aspects of social life and building SC. Broadband Internet can enhance distinct interactions which provide valuable links and opportunities. For example the relationships leveraged through weak ties can be very helpful on finding jobs or for obtaining information and knowledge using Internet forum. So that Broadband Internet plays an important role in facilitating two way communication or interaction and the synchronous and asynchronous characteristics of Broadband Internet facilitate brief interactions and multitasking, i.e. doing other things while interacting with different ties at the same time (Resnick 2001).

Since social capital is about networks, and Broadband Internet plays important role in connecting family, friends and community, then several questions arise concerning the relationship between the Broadband Internet and social capital. For example, due to its low cost and ubiquity of usages, Broadband Internet creates the opportunity of constant social communication i.e. connectivity and supports personal ties and connecting with larger communities with similar interest and share information (Wellman et al. 2003). The communication functions of Broadband Internet may make social interaction more convenient and efficient and help interpersonal exchange by desynchronizing in time and space (Penard and Poussing 2010). Furthermore, the information function of Broadband Internet facilitates the acquisition of information about places, times of social events, politics and civic initiatives. It reduces transaction costs of travel and even helps for individuals' to find out about opportunities for preferred social volunteer engagement and jobs (Bauernschuster et al. 2014).

Hence the literature supports the following hypothesis H1: Broadband connectivity builds and maintains social capital (bonding and bridging capital) in rural communities.

## 2.4  Social networking Sites

The growth of social networks online since their mainstream emergence in the last 10 years has been both rapid and dramatic, changing the purpose and the functionality of the Internet. From general chit-chat to propagating breaking news, from scheduling a date to following election results or coordinating a disaster response, from gentle humour to serious research, SNS are now used for a host of different reasons by various user communities (ITU 2014). SNS are websites that allow individuals to construct a public or semi-public profile within a bounded system; articulate a list of other users with whom they share a connection; and view and traverse their list of connections and those made by others within the system (Boyd and Ellison 2008). SNS such as Facebook, Twitter, LinkedIn and Instagram are types of online virtual communities, which are the most widely known and used these days (Hu and Ma 2010). Facebook has over 850 million daily active users around worldwide (Facebook 2015). Facebook is widely used for connecting family, friends, and friends of friends through its online social network. Twitter is a microblogging platform and people usually follow twits to get immediate updates on recent news and current affairs. Twitter has 316 million active users (Twitter 2015). Similiarly, Instagram has 400 milllion active users where users express their thoughts and status by sharing pictures or photos (Instagram 2015). More recently Linkedin has become prominent amongst professional people and has more than 380 million active users (LinkedIn 2015).

SNSs has become important tools for managing relationships with a large network of people who provide social support and serve as channels for useful information and other resources. Some of the most common and important features of social activities which can be done by using a SNS are direct communication with individuals and friends, passive consumption of social news and broadcasting communication with everyone in networks (Burke et al. 2011). Direct communication that is of personal nature is most likely to happen with one-to-one exchange of information. This is much more like to be done using emails and instant messages which are now supported by a number of SNS through messages, wall posts and chat. In SNS such as Facebook, members participate in social activities by disclosing information about themselves on their profiles or commenting on friends' pages. In addition, there are mechanisms to facilitate building and maintaining close relations by using a 'Like' button, comments and photo tags (Burke et al. 2011). In each of these actions, one user singles out another user, signaling that their relationship is very good and close. Thus, direct communication is likely to be useful for maintaining relationships with existing ties and encouraging the building of new ones and has potential to improve bonding social capital. In addition to informal and frequent communication and connection, increased the strength of one's social ties.

## 2.5  Social Networking Sites building Social Capital in Rural Communities

According to ABS (2014B) the growth in Internet access and usage in rural and regional Australian communities from 38% in 1998 to 79% in 2013 presents a significant opportunity for building social capital in these communities (Warburton et al. 2013). Boase et al. (2006) and Stern (2008) suggest that the Internet can be used to connect communities to each other and support the positive relationships in community engagement. Others, Steinfield et al. (2009), Stern and Dillman (2006) and Valenzuela et al. (2009) support this notion within and outside an organization and society and show how organizations can also engage with communities using SNS and the Internet. SNS can positively influence and build social capital in rural communities by fostering new avenues for communication and voluntary engagement (Stern and Adams 2010; Valenzuela et al. 2009) that might otherwise been difficult to achieve due to the tyranny of distance in rural areas. Using SNSs can lead to increased contacts and the broadening of social networks for rural communities. These connections may results increase in bonding social capital as users might be in position to provide emotional supports whenever needed (Boyd and Ellison 2008). Moreover, bonding can be valuable for community members to band together in groups and networks and support their collective needs.

Applying this argument to the field of online communication, connection and engagement suggests that bonding social capital would be stronger and enhanced in rural communities through the use of SNSs.

Hence the literature supports the following hypothesis H2: Households in rural communities with broadband access using SNSs will have higher levels of bonding capital than households in rural communities with broadband access not using-SNSs.

In contrast, broadcasting communication and passive consumption of social news and undirected messages are one of the novel features of SNS where one reads and others update. Some of the features such as News Feeds are a general broadcast which allow sharing of status updates, links, photos, public interactions between friends and friends of friends. News Feeds can be easily viewed and shared and contain information such as profiles pages, photos and comments if the SNS users have not modified their privacy settings (Burke et al. 2011). SNS is more suitable for informal communication between weak ties (Zhao and Rosson, 2009) but many users start using SNS for formal communication as well. Such as political leaders who use SNS to forward their opinions or beliefs to the general public (Bronstein and Aharony, 2015). SNSs are also increasingly used for sending invitations or to promote events and functions (Rebelo and Alturas, 2011). SNS allows individuals to access information such as events in communities and rising alarm and help during natural disaster (Vieweg et al., 2010). More importantly SNS create opportunities such as jobs which shared by network friends, to be able to get references that are otherwise unavailable. The overall improvement of individuals' well-being and quality of life are by-products of SC (Burke et al. 2011; Valenzuela et al. 2009). Several empirical studies highlight that bridging SC, with the associated benefits of enhanced social participation and social inclusion, can counter growing concerns of social isolation (Burke et al., 2011; Gray et al., 2006; Steinfield et al., 2008). This has particular importance in rural communities because the affordances of SNSs are well-suited to maintaining these ties cheaply and easily. In particular, bridging social capital might be augmented by social network sites such as Facebook because they enable users to create and maintain larger, diffuse networks of relationships from which they could potentially draw resources (Donath and Boyd, 2004, Resnick, 2001 and Wellman et al., 2001). Hence the literature supports the following hypothesis:

H3: Households in rural communities with broadband access using SNSs will have higher levels of bridging capital than households in rural communities with broadband access not using-SNSs.

Figure 1 provides a conceptual model of the hypothesised relationships that were tested and reported on in this paper.

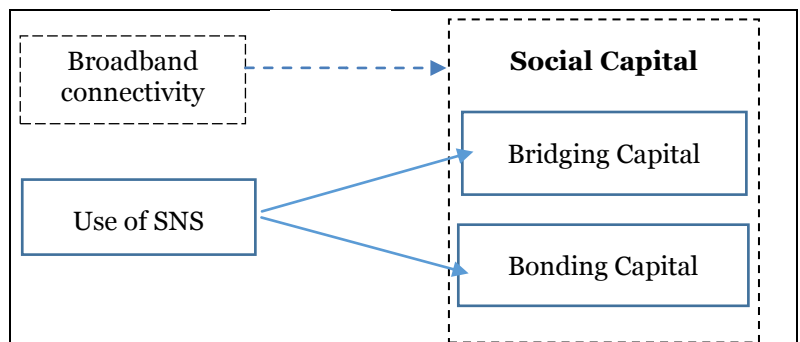

*Figure 1: Broadband connectivity and the use of Social networking sites builds and maintains Social capital (Bonding Capital and Bridging capital) in Rural Communities*

## 3  METHODOLOGY

This research employed a large scale survey in the Western Downs Region (WDR) on households' experiences with their adoption and use of Broadband Internet including their use of SNS. The survey sample population of 1,500 households was randomly selected from the different population centres in WDR, Queensland, Australia using a stratified sampling approach (Explorable, 2009 ). The WDR covers a land area of 38,039 square kilometers and had a population of 32,872 (as at 2012) making it 20th largest council in Queensland in terms of area. The population of the WDR is concentrated in the three largest towns, Dalby, Chinchilla and Miles with the rest of population dispersed across a number of smaller towns and rural districts. We used the RRMA (Rural, Remote and Metropolitan Areas) classification for this research (AIHW 2004). The population localities across WDR are classified as rural or remote using the RRMA classification because the urban centres within each of the population localities fall within the ranges of 10,000 to 24,999, and 5000 to 9,999 (rural urban centres), 100 to 4999 (remote urban centres). This natural distribution of population in WDR across three larger towns, a number of small towns and rural districts makes it a representative sample of rural Australia and an ideal setting for studying issues concerning the adoption and use of a technology such as Broadband Internet and SNS to build and maintain social capital with close ties and weak ties in rural communities.

The survey instrument was developed from a number of previous survey instruments including an instrument for measuring SC in the context of social networking sites (Ellison et al. 2007; Neves 2013; Williams 2006). Using survey instruments which have been previously proven to be valid and reliability increases the validity and reliability of survey results (Straub et al. 2004). The survey instrument was pretested with a number of academics experienced with survey research and the adoption and use of a technology such as Broadband Internet. The full survey was piloted with a number of households who reside in WDR before conducting the main data collection phase. The survey instrument is available on request from the authors and achieved a response rate of about 20 Percent (302 completed and usable survey responses shown in Table 1). The survey was distributed in person to randomly selected households using a stratified sampling method to ensure that the 1500 surveys were distributed to a representative sample size across population localities in WDR. The targeted respondent was the major decision maker in each household.

| **Location in WDR** | **Household Responses** | **Combined district percentage** |
|---|---|---|
| Live in Dalby | 126 (42%) | Dalby district combined 49% |
| Nearest to Dalby | 21 (7%) | |
| Live in Chinchilla | 29 (10%) | Chinchilla district combined 13% |
| Nearest to Chinchilla | 10 (3%) | |
| Live in Miles | 13 (4%) | Miles district combined 8% |
| Nearest to Miles | 10 (3%) | |
| Live in Jandowae | 15 (5%) | Jandowae district combined 7% |
| Nearest to Jandowae | 7 (2%) | |
| Live in Tara | 10 (3%) | Tara district combined 7% |
| Nearest to Tara | 12 (4%) | |
| Live in Wandoan | 14 (5%) | Wandoan district combined 6% |
| Nearest to Wandoan | 4 (1%) | |
| Other place (please specify) | 31 (10%) | Other places in WDR 10% |
| Total | 302 (100%) | |

*Table 1. Geographical distribution of respondent households in Western Downs Region survey*

The survey data was analysed using the statistical data analysis software SPSS and the structural equation modelling software smartPLS. In providing responses to the survey, the respondents were asked to indicate how much they agreed or disagreed on a seven-point Likert scale with series of statements that tapped into various dimensions that have previously been associated with a number of constructs including social capital. In order to analyse the data collected on social capital, first we conducted an exploratory factor analysis, using principal components with varimax rotation to confirm that the factors underpinning the measurement of social capital did not exhibit common method bias. A confirmatory factor analysis was used to assess the reliability and factorial validity of two dimensions measuring social capital (Bonding Capital, Bridging Capital) (Byrne 2001; Byrne 2013; Hair et al. 2010) in the context of broadband connectivity. We used independent samples T-Tests to determine if there were differences in the two dimensions of SC (Bonding Capital, Bridging Capital) between SNS users and non-SNS users.

# 4 RESULTS AND DISCUSSION

## 4.1 Demographics of households survey respondents

For survey respondents, the lowest representation in age categories were 18- 24 years (3%) and over 75 years (7%). The age category of 45- 54 years represented the largest group of respondents (28%), followed by 35-44 years (21.2%). The age categories 25- 34 and 55-64 years represented 18% and 16% respectively of respondents. A total of 202 (67.1%) male and 94 (31.2%) female respondents completed and returned the questionnaire. Almost half of survey respondents 144 (47.8%) are households classified as a couple or family with children at home, followed by 63 (20.9%) households classified as a couple or family with children not living at home. Ninety-two (31%) respondents' annual household income was between $20,000 and $59,999. A total 118 (39.3%) respondents had an annual household income in the middle range of between $60,000 and $120,000. In comparison, 57 (18.9%) respondents had a top end annual household income over $120,000. The highest number of respondents 143 (47.7%) possessed a secondary school education level followed by 117 (40.8%) respondents who had a higher education level of Diploma, Undergraduate or Post Graduate degree. Only 24 (8.4%) had done some training courses and less than 4% have only primary education.

## 4.2 Results of confirmatory factor analysis

As shown in Table 2, the survey items measuring social capital were confirmed as loading on two distinct dimensions of social capital – bonding and bridging social capital in the context of broadband connectivity.

The reliability and validity of the measurement of the constructs for SC in the context of broadband connectivity were assessed using a confirmatory factor analysis which determined whether the underlying manifest variables accurately reflect and measure their constructs. An assessment of the measurement model includes the reliability, convergent validity and discriminant validity are presented in Table 2. The composite reliability (CR), Cronbach alpha and average variance extracted (AVE) for each construct were used to confirm the reliability of all constructs (Byrne 2001; Hair Jr et al. 2006; Holmes-Smith 2005; Holmes-Smith 2011). Reliability and convergent validity were interpreted using 0.7 level, which has been widely suggested as the benchmark for moderate reliability. The CR and AVE values for constructs (bonding capital and bridging capital) exceeded the minimum acceptable values of 0.7 and 0.5 respectively (Byrne 2001; Hair et al. 2010; Holmes-Smith 2005), indicating a good reliability level and subsequently yielding very consistent results. To confirm the discriminant validity of constructs, we assessed the convergent validity by evaluating the constructs in terms of AVE. This should be greater than the variance shared between the other constructs and factor loadings and cross loading of the first order and second order model latent variables items which should be 0.5. All of constructs and factor loadings met the minimum acceptable values as shown in Table 2 and indicating good reliability and validity (Hair et al. 2010).

Common variance bias is a major systematic contributor to measurement error in survey research (Bagozzi and Yi 1991). To test for the extent of bias caused by common methods variance (CMV), Harman's single factor test was conducted using an exploratory factor analysis in IBM SPSS 22 (Podsakoff et al. 2003). As more than one factor emerged from an exploratory factor analysis to explain the total variance in the factor analysis, we can infer that common methods bias in this case is not high. Hence, common methods bias has been shown to have minimal effect and is not considered to be a concern in this study.

The results of the confirmatory factor analysis show SC and its dimensions of bonding and bridging capital provide strong support for the notion that SC can be built and maintained online as well as face to face. It would also appear that there is stronger support for building and maintaining bridging capital online through broadband connectivity than bonding capital.

| Social Capital | | Item statements | Mean | SD | Factor Loading | Cronbach's alpha | CR | AVE | R Square | H1 and H2 Beta Coefficients |
|---|---|---|---|---|---|---|---|---|---|---|
| **Bonding Capital** | | | | | | 0.85 | 0.91 | 0.52 | 0.73 | 0.85 (6.16 ***) |
| | BO_C1 | There are several people online my household trust to help solve our problems. | 4.04 | 1.77 | 0.73 | | | | | |
| | BO_C2 | There is someone online my household can turn to for advice about making very important decisions. | 3.90 | 1.79 | 0.73 | | | | | |
| | BO_C4 | When someone in my household feels lonely, there are several people online we can talk to. | 3.97 | 1.89 | 0.78 | | | | | |
| | BO_C5 | If my household needed an emergency loan of $500, we know someone online we can turn to. | 3.14 | 2.01 | 0.73 | | | | | |
| | BO_C6 | The people, my household interact with online would put their reputation on the line for my household. | 3.46 | 1.78 | 0.75 | | | | | |
| | BO_C7 | The people my household interact with online would be good job references for my household. | 3.79 | 1.78 | 0.81 | | | | | |
| | BO_C8 | The people my household interact with online would share their last dollar with my household. | 3.36 | 1.76 | 0.78 | | | | | |
| | BO_C9 reversed | My household does not know any people online well enough to get them to do anything important. | 3.93 | 2.01 | 0.65 | | | | | |
| | BO_C10 | There are several people online my household trust to help solve our problems. | 4.02 | 1.79 | 0.76 | | | | | |
| **Bridging Capital** | | | | | | 0.95 | 0.96 | 0.70 | 0.86 | 0.93 (7.49 ***) |
| | BR_C1 | Interacting with people online makes my household interested in things that happen outside of our town. | 4.91 | 1.70 | 0.87 | | | | | |
| | BR_C2 | Interacting with people online makes my household want to try new things. | 4.70 | 1.68 | 0.88 | | | | | |
| | BR_C3 | Interacting with people online makes my household interested in what people unlike my household are thinking. | 4.51 | 1.65 | 0.89 | | | | | |
| | BR_C4 | Talking with people online makes my household curious about other places in the world. | 4.82 | 1.70 | 0.88 | | | | | |
| | BR_C5 | Interacting with people online makes my household feels like part of a larger community. | 4.72 | 1.70 | 0.91 | | | | | |
| | BR_C6 | Interacting with people online makes my household feels connected to the bigger picture. | 4.80 | 1.69 | 0.91 | | | | | |
| | BR_C7 | Interacting with people online reminds my household that everyone in the world is connected. | 5.03 | 1.54 | 0.82 | | | | | |
| | BR_C8 | My household is willing to spend time to support general online community activities. | 4.36 | 1.58 | 0.85 | | | | | |
| | BR_C9 | Interacting with people online gives my household new people to talk to. | 4.41 | 1.67 | 0.85 | | | | | |

Legend BO_C = Bonding Capital; BR_C = Bridging Capital

*Table 2. Assessment of reliability and validity of social capital items and construct in the context of broadband connectivity*

In second part of the data analysis presented in this paper we sought to determine whether the use of SNSs resulted in higher levels of social capital for survey respondents given almost all (> 95%) of the survey respondents had broadband internet connectivity (there were a very small number of households still using dial-up). Table 3 shows the different types of SNS sites being used by households in the survey responses.

| Types of SNS | Count | Percentage |
| --- | --- | --- |
| Facebook | 214 | 71.6 |
| Instagram | 44 | 14.7 |
| LinkedIn | 35 | 11.7 |
| Twitter | 21 | 7.0 |

*Table 3. Type of social networking sites used to communicate by survey respondents*

Table 3 shows that that Facebook is most widely used SNS among the survey respondents. Seventy two percent of survey respondents are using Facebook followed by Instagram (14.7%), LinkedIn (11.7%) and Twitter (7%). We ran independent sample T- Tests on the survey responses to identify whether there is significant differences between the SNS users and Non–SNS users in terms of their ratings of SC (Bonding, Bridging) in rural and regional communities. Data is the mean ± standard deviation unless otherwise stated for the following independent samples T-Tests. For Bonding capital survey respondents were asked about whether through using Broadband Internet that their online communication with their close ties built trust in other words bonding capital. An independent-samples T-Test was run to determine if there were differences in the overall average rating of bonding capital between SNS users and Non-SNS users. There were 222 SNS users and 79 non-SNS users. There were no outliers in the data, as assessed by inspection of a boxplot and histogram. Bonding capital for both SNS users and non- users was normally distributed, as assessed by Shapiro-Wilk's test ($p > .05$), and there was homogeneity of variances, as assessed by Levene's test for equality of variances ($p = .080$). Bonding capital overall in an online context was perceived to be higher for SNS users (M=3.93, SD=1.34) than Non-SNS users (M=3.14, SD=1.07), with a statistically significant difference of 0.79 in the means of SNS users compared to Non-SNS users (95% CI, 0.46 to 1.22), $t(299) = 4.717$, $p = .000$, and a medium size effect (d) =0.62. These results suggest that SNS users have more trust and engagement and in other words higher levels of bonding capital with their close ties using Broadband Internet rather than non-SNS users. Our findings suggest SNSs may be assisting households with access to Broadband services within rural communities to build and maintain bonding capital.

For Bridging capital survey respondents were asked about whether through using Broadband Internet that their online communication with weak ties in outside communities was strengthened. An independent sample T-Test was ran to determine if there were differences in Bridging capital between SNS users and Non-SNS users. There were 215 SNS users and 60 non-SNS users. There were no outliers in the data, as assessed by inspection of a boxplot and histogram. Bridging capital for each level of SNS users and non- users were normally distributed, as assessed by Shapiro-Wilk's test ($p > .05$), and there was homogeneity of variances, as assessed by Levene's test for equality of variances ($p = .290$). There was a significant difference in the mean scores of Bridging capital for SNS users (M=5.15, SD=1.11) compared to non-user (M=4.44, SD=1.04) a statistically significance difference, of 0.71, (95% CI, 0.40, 1.02), $t(273) = 4.485$, $p < 0$ and a medium size effect (d) =0.65. These results suggest that households in rural communities that are SNS users using Broadband Internet are more engaged with weak ties in outside communities. In other words, SNS users would appear to have higher levels of bridging capital than Non-SNS users in the survey responses. The results of the analysis along with the hypotheses are summarized in Table 4.

| Hypotheses | Result |
| --- | --- |
| H1: Broadband connectivity builds and maintains bonding and bridging social capital in rural communities | Supported |
| H2: Households in rural communities with broadband access using SNSs will have higher levels of bonding capital than households in rural communities with broadband access not using-SNSs. | Supported |
| H3: Households in rural communities with broadband access using SNSs | Supported |

will have higher levels of bridging capital than households in rural communities with broadband access not using-SNSs.

*Table 4. Summary of the results of Hypothesis testing for Broadband Connectivity and Social networking sites build and maintain Social capital (Bonding Capital and Bridging capital) in Rural Communities*

## 5　DISCUSSION OF KEY FINDINGS

In this study of Broadband connectivity in rural communities, bridging capital or weak ties (which helps individuals to communicate with outside or distance communities or people) is identified as the most significant factor in building social capital in rural communities such as in WDR. For bridging capital, the following survey item "interacting with people online reminds my household that everyone in the world is connected" was considered the most important item by the survey respondents (M= 5.03 and SD=1.54,) and in contrast the following survey item "my household is willing to spend time to support general online community activities", (M=4.36 and SD=1.58, on a seven point likert scale) was rated least important item for bridging capital. In contrast, bonding capital has relatively lower mean valves compared to bridging capital in the survey responses. This might be because people often like proximity and face to face interactions are important in the maintenance of community ties in rural communities. SNS are just another communication tool for maintaining these close ties and bonding capital and an extension but not a replacement for face to face communication.

Similar to Ellison et al. (2007) this study found a strong association between use of SNS and social capital, with the strongest relationship with SNS being with bridging capital. Therefore the evidence presented in this paper tentatively does suggest that SNSs provide a mechanism for households in rural communities to build and maintain social in both its forms – bonding capital and bridging capital.

Our research findings suggest that broadband connectivity and SNS are more efficient and effective in building and maintaining bridging capital then bonding capital.  Results overall support the notion that broadband Internet connectivity and use of SNS helps to build and maintain networks in rural communities and increase their social well-being, civic engagement and trust (Neves 2013; Valenzuela et al. 2009).

## 6　CONCLUSION

In this research, we use an instrument to confirm the measurement of social capital and its two key dimensions of bonding capital and bridging capital in the context of Broadband connectivity. We also considered the use of SNSs as a facilitating mechanism for building and maintaining social capital for households in rural communities that have broadband connectivity. We provided evidence that suggests the use of SNSs by households in rural communities plays a vital role in connecting with them with their strong ties and weak ties and generally building social well-being in these rural communities. Findings of the study suggest that households in rural communities with broadband connectivity that are using SNS have higher levels of SC compared to non-SNS users with broadband connectivity. Our findings suggest that those who are communicating both online and offline have better links and networks in their rural communities than those don't make use of broadband connectivity and SNS to communicate.

There are significant implications of Broadband Internet connectivity and use of SNS, to connect online and offline within communities' strong and weak ties. Our results suggest that broadband connectivity contributes significantly to building and maintaining bridging capital but marginally to building and maintaining bonding capital. However, people in rural communities are using SNSs such as Facebook to keep connected with family, relative and friends who have often moved to urban areas. Results reported in this paper also indicate that rural communities are using different kinds of SNS such as Facebook, Instagram, LinkedIn and Twitter to communicate with outside communities. This is the beauty of SNS that it can be used to create and manage rural community groups and events. Rural communities can participate and support each other both online and offline using SNS, which ultimately helps to build social capital in rural communities. This study shows that network effect among rural communities is really important for building social capital which has practical benefits and could help in many ways when support is needed from a rural community. Local government could utilize SNS as an information and connecting medium for communicating with rural communities in emergency situations. Rural communities could able to access inclusive social opportunities and connectedness through their participation in SNS. Here we show preliminary results of possibilities of broadband connectivity and the use of SNS building and maintaining SC in rural communities. However, further investigation needs to be undertaken to examine in more details the role of Broadband connectivity and social media in general in building SC in rural communities. Future research using longitudinal studies of broadband connectivity, digital literacy and digital competency could establish if there is any relationship among these and how these contribute to building and maintaining SC in rural communities such as WDR.

## Acknowledgements